\documentclass[journal]{IEEEtran}
\usepackage{cite}
\usepackage{graphicx}
\usepackage{amsmath,bm,mathtools,siunitx,amssymb}
\DeclareSIUnit\sq{sq}
\usepackage{array}
\usepackage{float,stfloats,comment}
\usepackage{colortbl,caption,tabularray, tabularx}
\UseTblrLibrary{booktabs}
\usepackage{url}
\usepackage[hidelinks]{hyperref}
\usepackage{soul}
\usepackage[table]{xcolor}
\definecolor{laser}{HTML}{bd0c0c}
\definecolor{signal}{HTML}{bd0c0c}
\definecolor{waveguide}{HTML}{7500ff}
\definecolor{modulation}{HTML}{447821}
\definecolor{GND}{HTML}{000000}
\definecolor{reference}{HTML}{0044aa}
\definecolor{current}{HTML}{d45500}
\definecolor{modulation}{HTML}{03a27a}
\definecolor{TitleBlue}{HTML}{0044aa}
\definecolor{RowBlue}{HTML}{E6F0FF}
\DeclareRobustCommand*{\IEEEauthorrefmark}[1]{%
  \raisebox{0pt}[0pt][0pt]{\textsuperscript{\footnotesize #1}}%
}

\begin{document}

\title{A single-step lithography process for reconfigurable SiN photonics with TiN heaters and Al interconnects}

\author{\IEEEauthorblockN{
        Leonardo Limongi\IEEEauthorrefmark{1,2},
        Rachele Favaretto\IEEEauthorrefmark{1,3,4},
        Lorenzo Baldessarini\IEEEauthorrefmark{1},
        Martino Bernard\IEEEauthorrefmark{1},
        Yong Kwon\IEEEauthorrefmark{1,3},
        Gioele~Piccoli\IEEEauthorrefmark{1},
        Alina Samusenko\IEEEauthorrefmark{1},
        Georg Pucker\IEEEauthorrefmark{1},
        Mher Ghulinyan\IEEEauthorrefmark{1}
        }\\
        \IEEEauthorblockA{
        \IEEEauthorrefmark{1}Center for Sensors and Devices, Fondazione Bruno Kessler, Via Sommarive 18, 38123 Trento, Italy\\
        \IEEEauthorrefmark{2}Department of Industrial Engineering, University of Trento, Via Sommarive 9, 38123 Trento, Italy\\
        \IEEEauthorrefmark{3}Department of Physics, University of Trento, Via Sommarive 14, 38123 Trento, Italy\\
        \IEEEauthorrefmark{4}Femtorays Technologies (FTH s.r.l.), Via Sommarive 18, 38123 Trento, Italy\\
        }
        }

\maketitle

\begin{abstract}
Thermo-optic phase shifters are key building blocks in Silicon and Silicon Nitride-based reconfigurable photonic integrated circuits. They enable manipulating the phase of an optical signal by means of electrically-driven heating of an optical waveguide. Conventional fabrication schemes typically require dedicated lithographic steps to separately define the resistive heaters, the current transmission lines, and the electrical contact pads. This increases the process complexity and slows the standard complementary metal-oxide-semiconductor (CMOS) fabrication flows. In this work, we present a single-step lithographic process for the realization of Titanium Nitride thermo-optic phase shifters and Aluminum interconnects integrated on a Silicon Nitride photonic platform. A detailed electro-optical characterization, performed on two platforms operating at 810 nm and 1550 nm, revealed $\pi$-shift powers of 92 $\pm$ 2 mW and 120 $\pm$ 10 mW, respectively. Alongside, modulation bandwidths of 8.5 $\pm$ 0.3 kHz and 3.83 $\pm$ 0.03 kHz were extracted from combined frequency- and time-domain analyses. Our results demonstrate that the proposed single-step lithographic metal definition process represents a robust, viable and cost-efficient route towards CMOS-compatible reconfigurable Silicon Nitride photonics.\newline
\end{abstract}


\section{Introduction}\label{sec:Introduction}
Phase shifters are fundamental components in optical experiments, utilized for light modulation by altering the optical phase to induce a controlled interference. In silicon photonics, these devices are typically integrated into waveguides within photonic integrated circuits (PICs) to manipulate the optical path of the light\cite{xiang2022silicon,pelucchi2022potential,aharonovich2026programmable}. A common implementation is the thermo-optic phase shifter (TOPS or \textit{heater} as well), which utilizes a resistive material deposited onto the waveguide to dissipate heat via Joule heating\cite{densmore_compact_2009, smith_phase-controlled_2009}. The released heat alters the refractive index of the waveguide material proportional to the local rise of temperature because of the thermo-optic effect. This generates a precise phase difference between the light signals propagating in heated and non-heated waveguides.

Since TOPS operation relies on electrical drive power and thermal gradients, managing these dependencies remains a challenging task\cite{harris_efficient_2014, jacques_optimization_2019, parra_silicon_2024}. Specifically, as the driving power increases to achieve larger phase shifts, heater components may suffer from permanent damage or structural failure if their critical resistance thresholds are exceeded. This risk is further compounded by inefficient heat localization, which often leads to thermal crosstalk, where leaked heat inadvertently affects adjacent optical components. To circumvent these thermal bottlenecks, phase shifters based on micro-electromechanical systems (MEMS) can be employed; these incorporate structural features such as air gaps or suspended membranes to provide superior thermal isolation, significantly reducing power consumption compared to conventional lateral heat diffusion methods\cite{sun_silicon_2022, sattari_silicon_2019}. However, MEMS-based phase shifters often require a complex post-fabrication process, such as the selective removal of sacrificial oxide layers to release suspended structures or the implementation of wafer-level hermetic sealing\cite{errando-herranz_mems_2020, quack_integrated_2023}. In contrast, CMOS foundries offer a more standard and straightforward TOPS fabrication flow\cite{siew_review_2021, liu_thermo-optic_2022}. Due to this advantage, TOPS are widely utilized in various fields, from programmable analog calculations \cite{bogaerts2020programmable} and quantum information processing \cite{aharonovich2026programmable,kwon_full_2024}, to microwave photonics \cite{zhuang2015programmable,marpaung2019integrated}, LIDAR \cite{derose2013electronically} and biosensing applications\cite{FavarettoAISEM}. 

\begin{figure*}
\centering
\begin{minipage}[b]{.5\textwidth}
\centering
\strut\vspace*{-\baselineskip}\newline\includegraphics[width=\linewidth]{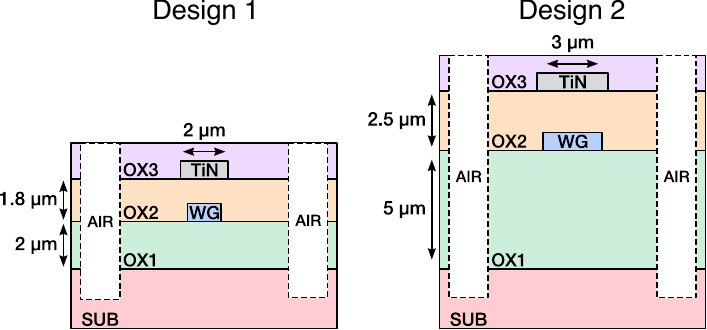}
\end{minipage}
\hfill%
\begin{minipage}[b]{.46\textwidth}
\centering
\captionof{table}{\textbf{Summary of key materials dimensions.} Symbols $\updownarrow$ and $\leftrightarrow$ indicate thickness and width respectively.}\label{tab:Material_stack}
\footnotesize
\begin{tblr}{
            colspec = {X[5,c,m]
                       X[10,c,m]
                       X[10,c,m]
                       X[10,c,m]},
            rowsep = 5pt,
            row{1} = {bg = TitleBlue, fg = white},
            row{2} = {bg = RowBlue},}
    \hline
    Design & OX1 $\updownarrow$ (\unit{\um}) & OX2 $\updownarrow$ (\unit{\um}) & TiN $\leftrightarrow$ (\unit{\um})\\
    \hline
    1 & 2.0 & 1.8 & 2.0 \\
    2 & 5.0 & 2.5 & 3.0 \\
    \hline
\end{tblr}
\strut\vspace*{-2\baselineskip+5pt}\newline
\end{minipage}
\caption{\textbf{Cross-sectional schematics of Design~1 and Design~2 (not to scale).} Both configurations utilize a common material stack consisting of a substrate (SUB), bottom oxide (OX1), waveguide core (WG), cladding oxide (OX2), a Titanium Nitride (TiN) layer, and an upper oxide cladding (OX3). Deep AIR trenches  provide lateral thermal isolation. Design~1 presents a more compact vertical profile with an OX1 of \qty{2}{\um}, OX2 of \qty{1.8}{\um}, and a narrower TIN width of \qty{2}{\um}. Design~2 features a thicker geometry with an OX1 of \qty{5}{\um}, OX2 of \qty{2.5}{\um}, and a TiN width of \qty{3}{\um}.}\label{fig:Material_stack}
\end{figure*}

To exploit these advantages while mitigating adverse electrical and thermal effects, the selection of an optimized resistive material is crucial. In silicon nitride (SiN) photonics, various materials have been investigated for heaters, including Tungsten (W)\cite{lee_controlled-not_2022}, Nichrome (NiCr)\cite{lin_power-efficient_2024}, Titanium (Ti)\cite{deleau_electro-thermo-optical_2026}, and Titanium Nitride (TiN)\cite{yong2022power, gao_comprehensive_2023, gebregiorgis_highly-efficient_2025}. 
In this work, we select TiN as the primary resistive material and demonstrate a TOPS geometry that incorporates Aluminum (Al) films outside the active heating regions, with a proposed optimized fabrication process. In this architecture, the Al film serves as a low-resistance conductive layer for interconnects and bond pads, ensuring efficient power delivery across the PIC. Notably, both materials are CMOS-compatible and we show how they can be deposited and etched in a single lithographic step, significantly simplifying the fabrication workflow. We provide a systematic characterization of the optical, electrical, and thermo-optical properties of these TiN-based TOPS integrated with SiN waveguides. Specifically, we evaluate such TOPS by defining their Safe Operating Area (SOA) to determine their current and thermal tolerances, the $\pi$-shift power ($P_\pi$), and modulation response time ($\tau$). The results of two designs, operating at 810 nm and 1550 nm, respectively, are presented, with their structural differences being summarized in Fig.~\ref{fig:Material_stack}.
The remainder of this paper is organized as follows: Section~\ref{sec:Multilayer_Structure} details the design principles and fabrication of Al-Ti-TiN-Ti multilayer, Section~\ref{sec:Characterization} describes the experimental characterization of the device performance, and Section~\ref{sec:Conclusion} provides concluding remarks.

\section{Al-Ti-TiN-Ti multilayer structure}\label{sec:Multilayer_Structure}

To fabricate a PIC which exploits thermo-optic effects two conditions must be met at the same time. On one hand, to impact the waveguide temperature, the thermal source (heater) must be placed as close as possible to the waveguide, while the thermal sink (substrate) should be far away on the opposite side. On the other hand, the waveguide should remain optically insulated from both the substrate and any lossy/high-index metal film of the electric circuit that drives the heat source.
Depending on the optical loss and thermo-optic efficiency expectations, an optimal thickness for the oxide layer separating the waveguide and the heat source can be found, for example via Finite Element Method (FEM) simulations both in the heat-transfer and optical-mode domains. 
It is worth noting that both conditions require a thick bottom cladding, that will separate the waveguide from the high-index silicon substrate, which also acts as the thermal bath for the circuit. Thus, when steady state phase control is considered, the bottom cladding is only limited by the fabrication constraints. 
However, when the dynamic modulation of the thermo-optic actuator is taken into account, the thermal resistance and the capacitance of the bottom cladding will limit the bandwidth at which the device could be operated. This suggests to limit the bottom cladding thickness to the minimum requirements for the optical loss towards the substrate. 

For the purpose of diversifying and expanding the specificity of the thermo-optical characterization of the demonstrated heaters, two different PIC designs, with target operation wavelengths of \qty{810}{\nm} and \qty{1550}{\nm}, were realized and measured. The two versions (hereafter, Design~1 and Design~2, respectively) differ for the thickness of the silicon oxide claddings and for the width of both the heater and the waveguide. While all the other design parameters are shared, the differences in the cladding thickness will induce differences in the heat-transfer characteristics. These aspects will be in the focus of the investigations covering the following sections.

To address such design trade-offs, the first subsection discusses the optimization study using FEM simulations, while the second subsection provides a complete description of the fabrication process.

\subsection{Optical Design}\label{sec:Designing}

\begin{figure}[t]
    \centering
    \includegraphics[width=0.95\linewidth]{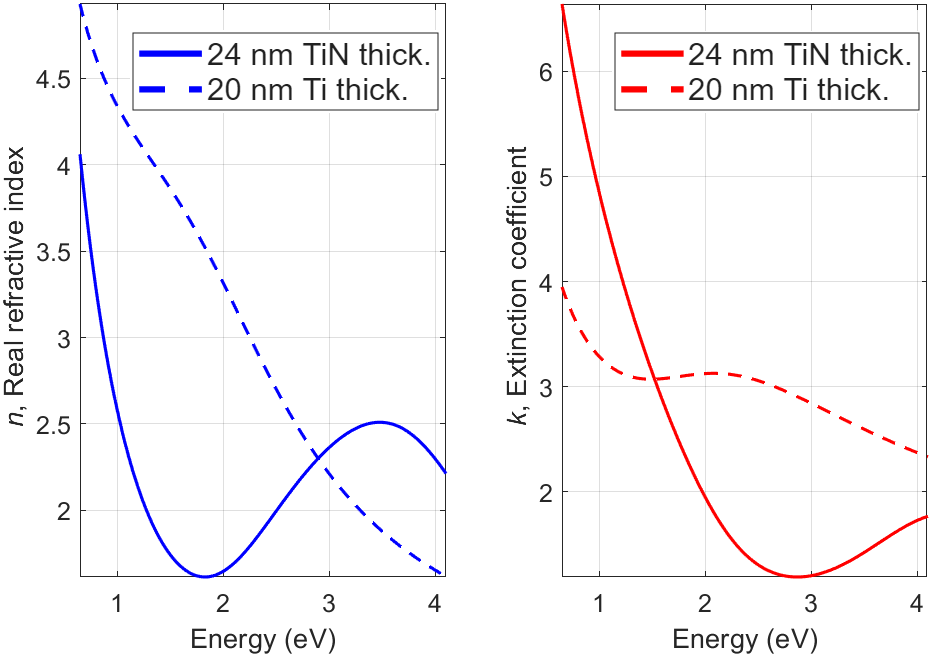}
    \caption{\textbf{Materials dispersion obtained from ellipsometric data.} The dispersions of the refractive index $n$ and the extinction coefficient $k$ of both \qty{24}{\nm} thick TiN (solid line) and \qty{20}{\nm} thick Ti (dashed line) films are reported.}
    \label{fig:Ti-TiN-Ti_Properties}
\end{figure}

The thermo-optical design requires an efficient power transfer from the electrical source to the waveguide in the form of heat.
We used FEM simulations in order to optimize the device geometry for the power transfer.

Regarding the bottom cladding (OX1), which ensures light isolation from the substrate, we chose from our FEM simulations to utilize a \qty{2.0}{\um} bottom cladding for Design~1 and a \qty{5.0}{\um} bottom cladding for Design~2. The reason for such difference is the mode dimensions at the two different operating wavelengths of Design~1 and Design~2.

To optimize the top cladding oxide (OX2) located between the waveguide and the heater, the knowledge of the refractive indices, $n(\omega)$ and extinction coefficients, $k(\omega)$ of TiN and Ti is essential. Figure~\ref{fig:Ti-TiN-Ti_Properties} shows the corresponding $n,k$ dispersions extracted form ellipsometric data of \qty{24}{\nm} thick TiN and \qty{20}{\nm} thick Ti films (refer to Sec.~\ref{sec:Fabrication} for a complete description of the fabrication process). Using these dispersions, on the basis of our simulation results, we chose to use an top oxide cladding of \qty{1.5}{\um} for Design~1 and \qty{2.5}{\um} for Design~2.

Furthermore, to achieve a better thermal confinement, lateral deep trenches (AIR) are considered in both designs. Notably, our optical simulations already consider a geometry with lateral trenches due to the finite width of the simulation window. Figure~\ref{fig:Material_stack} and Table~\ref{tab:Material_stack} summarize the main characteristics of both designs.

Regarding heaters and waveguide dimensions, instead, Design~1 features a waveguide of \qty{140}{\nm} $\times$ \qty{650}{\nm} (thickness $\times$ width) and utilizes a resistive strip width of \qty{2.0}{\um}, whereas Design~2 utilizes a \qty{350}{\nm} $\times$ \qty{1000}{\nm} waveguide and a resistive strip width of \qty{3.0}{\um}, in accordance to our standard technological baselines for devices operating at Near-Infrared (NIR) and Telecom spectral windows.

\subsection{Device fabrication}\label{sec:Fabrication}
\begin{figure*}[t]
    \centering
    \includegraphics[width=0.95\linewidth]{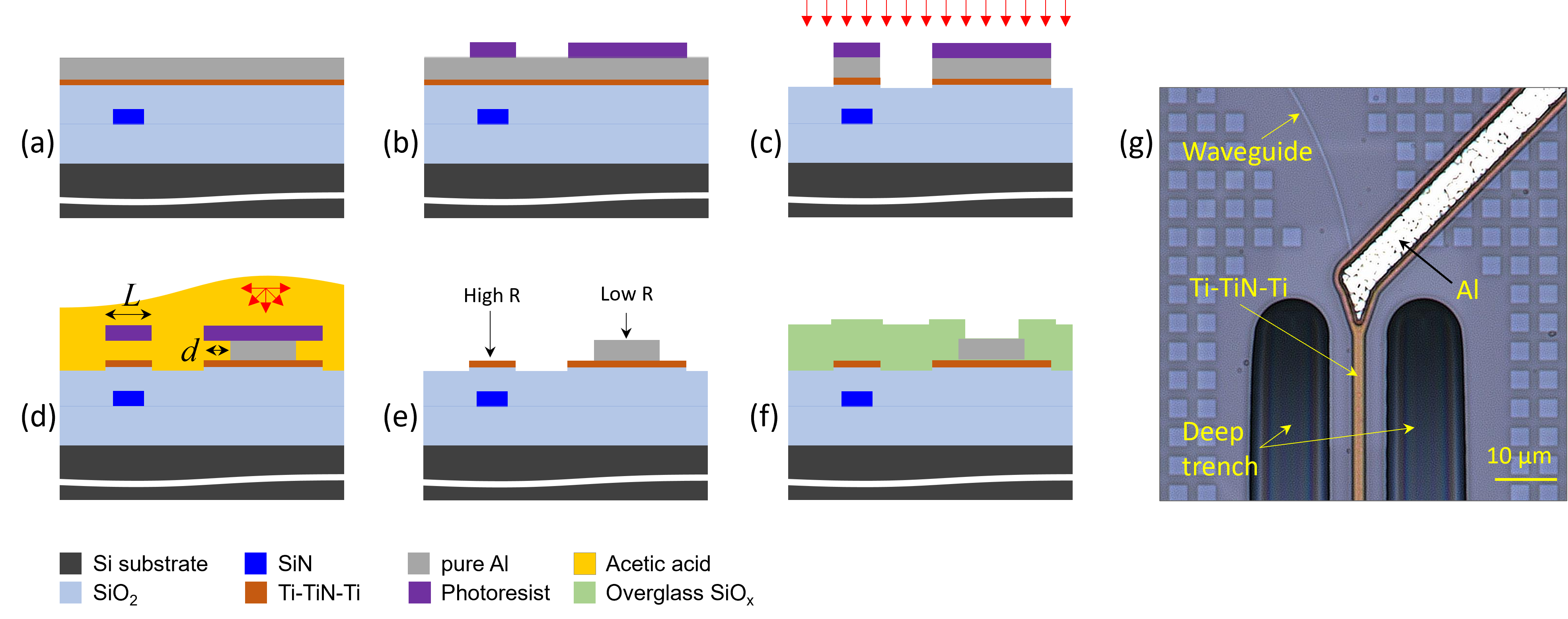}
    \caption{\textbf{A generic process flow of a single-step lithography metal definition for reconfigurable SiN PIC} \textbf{(a)} RF magnetron sputtering of the multi-metal Al-Ti-TiN-Ti stack. Intermediate 20~nm thick Ti films are used to promote adhesion at TiN/$\text{SiO}_2$ and TiN/Al interfaces. \textbf{(b)} The photoresist patterning of narrow phase shifters and low-resistivity current supply lines is \textbf{(c)} transferred to the full multi-metal stuck via anisotropic dry RIE. \textbf{(d)} A chemical selective isotropic wet etching removes completely the Al film from narrow metal patterns, \textbf{(e)} definition of the high-resistivity TiN phase shifters and low-resistivity Al-Ti-TiN-Ti metal lines. \textbf{(f)} A PECVD $\text{SiO}_x$ overglass film is deposited to protect the metal structures and eventually etched on large pads for electrical contacting. \textbf{(g)} An optical microscope image of a fabricated chip at the vicinity of the TOPS.}
    \label{fig:process_Al}
\end{figure*}

Our fabrication process for a generic reconfigurable PIC starts with the growth of a silica film by thermal oxidation of the silicon substrate at \qty{975}{\degreeCelsius} to form the bottom cladding. The silica film thickness can range from \qtyrange{2}{5}{\um} depending on the  spectral range of operation, from the Visible-NIR to the Third Telecom window. Then a silicon nitride waveguiding film is deposited using Low-pressure Chemical Vapor Deposition (LPCVD SiN). Next, the photonic circuits are defined by photoresist patterning using an i-line stepper lithography. The pattern is transferred to the SiN film using an anisotropic,  inductively-coupled plasma etch. After the removal of the residual photoresist, the wafer is covered with a doped silica film (Borophosphosilicate glass, BPSG) using LPCVD in order to form the top-cladding of the PIC. Specifically, the BPSG film thickness is \qty{1.8}{\um} for Design~1 and \qty{2.5}{\um} for Design~2, and after deposition it is reflowed in an N$_2$ ambient for \qty{90}{\minute} at \qty{925}{\degreeCelsius}. This planarizes partially the step topography of the SiN waveguides and allow for a good quality metal film deposition in the next processing step.

We use a physical vapor deposition (PVD) sputtering system to deposit the Al, Ti and TiN films. 
Prior to the realization of the complete Al-Ti-TiN-Ti multilayered stack, we separately prepared test substrates with thin, optically semitransparent TiN and Ti films of \qty{24}{\nm} and \qty{20}{\nm}, respectively, in order to characterize the optical constants through ellipsometric measurements (Fig.~\ref{fig:Ti-TiN-Ti_Properties}).
This allowed us to simulate the stack of materials for optical confinement, as previously discussed in Sec.\ref{sec:Designing}.

Continuing with the fabrication process, the wafers are covered with the Al-Ti-TiN-Ti multilayer in order to define ($i$) the high-resistivity TiN-based TOPS, ($ii)$ the low-resistivity current transmission lines and ($iii$) the electrical contact pads in a single lithographic patterning step (Fig.~\ref{fig:process_Al}a). We recall that, in our technology, the Al layer serves as the main material for the highly conductive electrical current lines, while TiN is used to realize the resistive TOPS. We also introduce two \qty{20}{\nm} thick Ti layers to guarantee a good adhesion of the TiN film to the bottom BPSG and to the top Aluminum films, respectively.     

The wafers are then lithographed with a photoresist pattern that contains all the above-listed metallic components (Fig.~\ref{fig:process_Al}b). After, the pattern is transferred from the photoresist to the multilayer in two consecutive etching steps: first, the Al-Ti-TiN-Ti stack is dry etched anisotropically down to the dielectric BPSG film using a chlorine-based reactive ion plasma etcher (Fig.~\ref{fig:process_Al}c). At the end of this etch, the wafer is transferred into a PAN etchant solution (76\% H$_3$PO$_4$ + 3\% HNO$_3$ + 3\% CH$_3$COOH + 18\% H$_2$O) to selectively etch the Al layer (Fig.~\ref{fig:process_Al}d). Being protected from top with the photoresist, the Al film is etched only laterally during this wet etching step. As a consequence, for an etching extension of $d$, the Al film is completely removed in all features with lateral dimensions $L\leq 2d$. In this way, the narrowest features in the lithographic pattern, which are the micro-resistors forming the TOPS, are freed completely from the top Al film, leaving only the high resistivity Ti-TiN-Ti multilayer structure. Instead, the remaining wider features, which are the current transmission lines and contact pads, still contain the highly conducting Al film (Fig.~\ref{fig:process_Al}e).

Successively, the metal pattern is protected by depositing from \qtyrange{1}{1.5}{\um} thick silica overglass film using PECVD. The overglass film is then removed on top of the contact pads via patterning and dry plasma etching (Fig.~\ref{fig:process_Al}f). In a final step,  the chip boundaries, the waveguide facets and trenches around the TOPS are defined by dry plasma etching of the full dielectric multilayer stack, completed with an additional  \qty{150}{\um} deep etch into the Si substrate through a Bosch process. This step facilitates butt-coupling between optical fibers and waveguides, while the deep trenches around TOPS help improve the efficiency of the thermo-optical effect on the SiN waveguides (Fig.~\ref{fig:process_Al}g).

\section{Device characterization}\label{sec:Characterization}

\begin{figure*}[t]
    \centering
    \includegraphics[width = \linewidth]{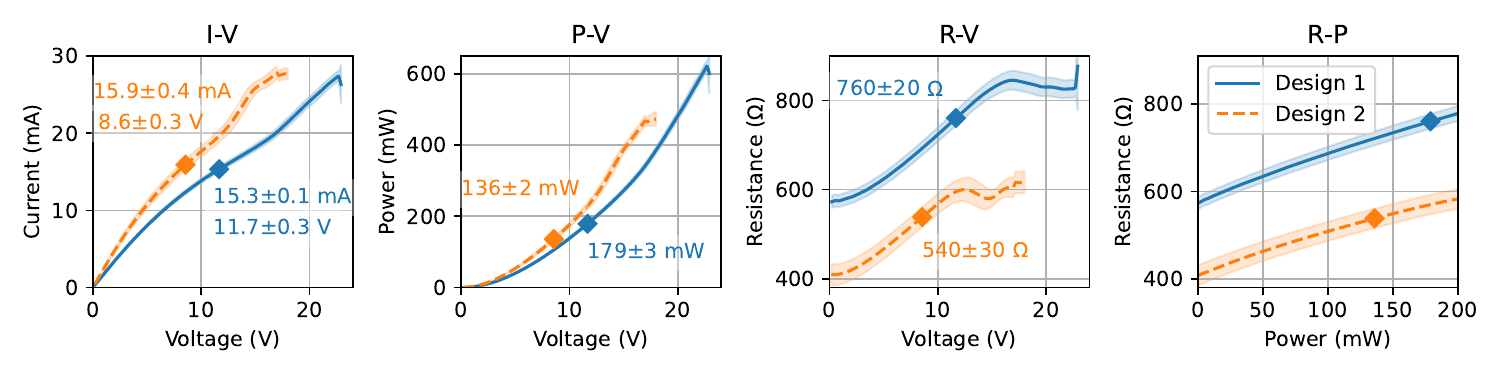}
    \caption{\textbf{Electrical characterization of the heater resistors for Design~1 (solid blue) and Design~2 (dashed orange):} $I$--$V$, $P$--$V$, $R$--$V$, and $R$--$P$ curves. Solid lines denote the mean response across the analyzed devices and shaded regions the corresponding standard deviation. The diamond markers denote the characteristic transition point identified from the $R$--$V$ curve and used to define the upper limit of the safe operating area. The corresponding mean values of voltage, current, dissipated power, and resistance are reported for each design.}
    \label{fig:IV_PV_RV_RP}
\end{figure*}

The characterization of the thermo-optic phase shifters was carried out in two stages. First, the electrical behavior of the heater resistors was investigated through destructive voltage sweeps on sacrificial devices in order to determine the safe operating area (SOA) in terms of voltage, current, and dissipated power (Section~\ref{sec:Heaters_Working_Range}). These limits were then used to define the operating conditions for all subsequent measurements.

In both chip designs, the heaters are integrated in a Mach--Zehnder interferometer (MZI), where the thermally induced refractive-index variation produces a differential phase shift between the two optical arms. The electro-optical response of the phase shifters was therefore characterized by extracting the phase modulation from the MZI transmission as a function of the applied electrical power (Section~\ref{sec:Electro_Optical_Characterization}). Both static and dynamic measurements were performed, yielding the $\pi$-shift power, modulation bandwidth, and thermal time constants.

\subsection{Electrical characterization and safe operating area}\label{sec:Heaters_Working_Range}
\begin{figure}[h]
    \centering
    \includegraphics[scale=1.1]{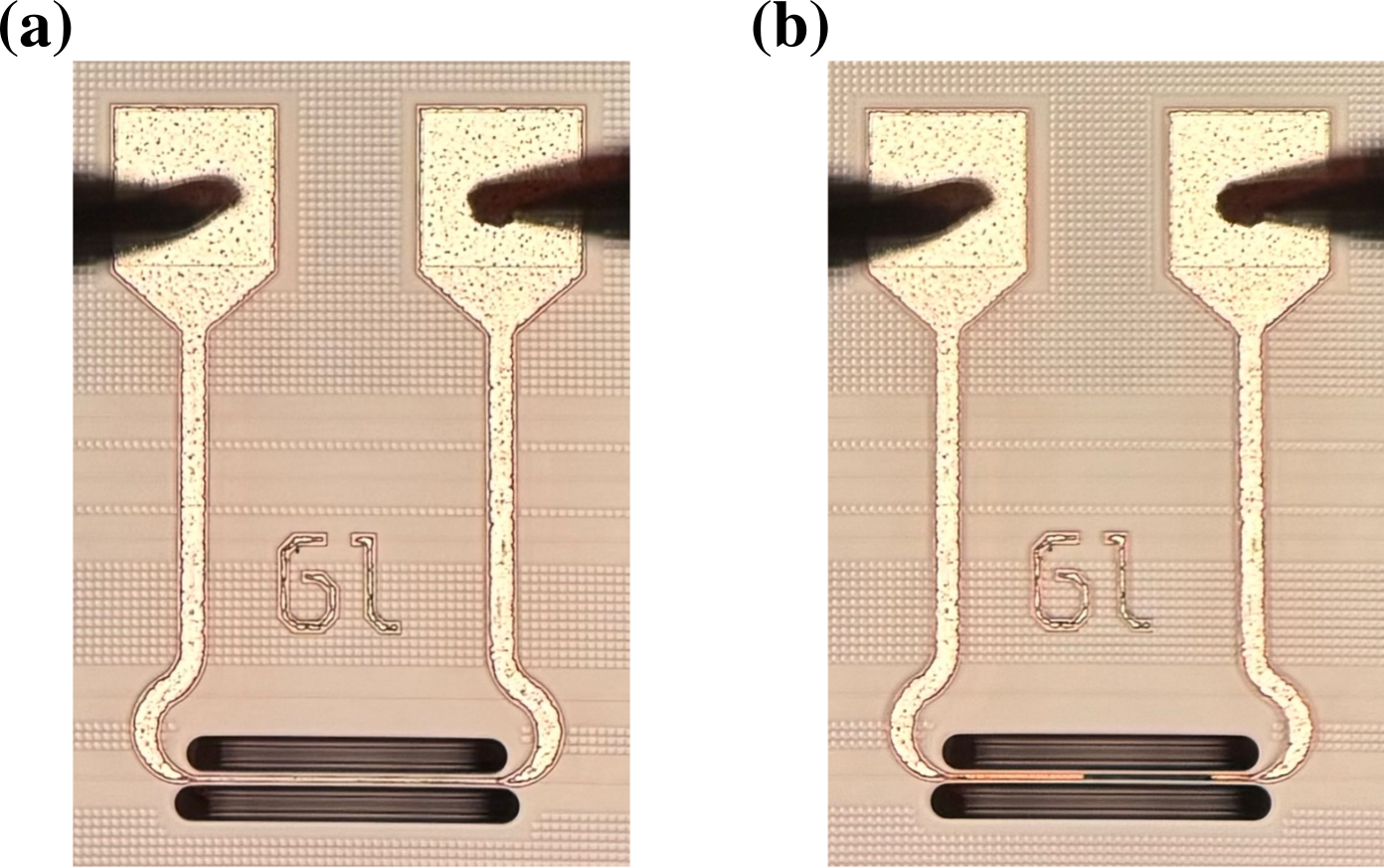}
    \caption{\textbf{Optical image of a resistor connected by tungsten probe tips.} \textbf{(a)} The intact resistor prior to testing. \textbf{(b)} The burnt resistor after failure induced by a voltage ramp, showing a visible open-circuit break in the trace.}
    \label{fig:Before_After_Failure}
\end{figure}

To establish safe operating conditions for the electro-optical measurements, the electrical breakdown of the heater resistors was characterized by means of voltage sweeps up to failure. A probe station equipped with two micro-positioned tungsten tips was used to contact the metal pads at the terminals of each heater, as shown in Fig.~\ref{fig:Before_After_Failure}. The electrical characterization was performed on 10 nominally identical heaters for each design.

A voltage sweep from \qtyrange{0}{25}{\volt} was applied in steps of \qty{0.1}{\volt}, while the resulting current $I$ was simultaneously recorded; the dissipated power $P$ and the instantaneous resistance $R$ were then derived from the acquired $(V, I)$ pairs via Ohm's law. The dissipated power was calculated as $P = VI$ from the voltage and current measured at the device pads. This value therefore represents the total electrical power delivered to the contacted heater structure, including any residual series resistance from the on-chip routing. Since the Al interconnects are designed to have a much lower resistance than the TiN heater strip, this contribution is expected to be small. The measured $I$--$V$, $P$--$V$, $R$--$V$, and $R$--$P$ characteristics are reported in Fig.~\ref{fig:IV_PV_RV_RP}, where the data for the Design~1 and Design~2 heaters are shown in solid-blue and dashed-orange lines, respectively. The reported curves represent the mean response, while the shaded regions indicate the standard deviation across the measured devices. Following these measurements, the resistors exhibited visible physical damage, as shown in Fig.~\ref{fig:Before_After_Failure}b.

The upper limits of the SOA were identified from the $R$--$V$ characteristics. For each device, the numerical derivative $dR/dV$ was computed and smoothed with a moving-average filter with 5 points to reduce spurious fluctuations. The characteristic transition point was then determined by locating the first peak in $\lvert dR/dV \rvert$, which corresponds to the first change in the slope of the resistance curve. This criterion identifies the onset of deviation from quasi-linear electrothermal behavior, which is interpreted as a precursor to irreversible device degradation.

The voltage at this transition point was taken as the upper boundary, and the corresponding current, resistance, and dissipated power were extracted from the same acquisition point. These quantities were averaged over the analyzed devices for both Design~1 and Design~2, and are summarized in Tab.~\ref{tab:Operating_Conditions} along with their standard deviations. Consequently, these SOA values represent upper operating bounds for all subsequent measurements.

The extracted SOA values demonstrate that the two designs fail at significantly different voltages and dissipated powers, which is consistent with their different electrical resistances. In contrast, the corresponding breakdown currents are remarkably similar. This suggests that the onset of irreversible degradation is primarily governed by current-carrying capability and the associated local Joule heating of the TiN heater, rather than by the applied voltage alone. The lower $V_\mathrm{max}$ and $P_\mathrm{max}$ values observed for Design~2 are therefore consistent with its lower resistance, which causes the critical current to be reached at a lower applied voltage. Additionally, the thicker oxide layer in Design~2 enhances thermal confinement, leading to higher local temperatures and contributing to an earlier onset of damage.

\begin{figure*}[t!]
    \centering
    \includegraphics[width = \linewidth]{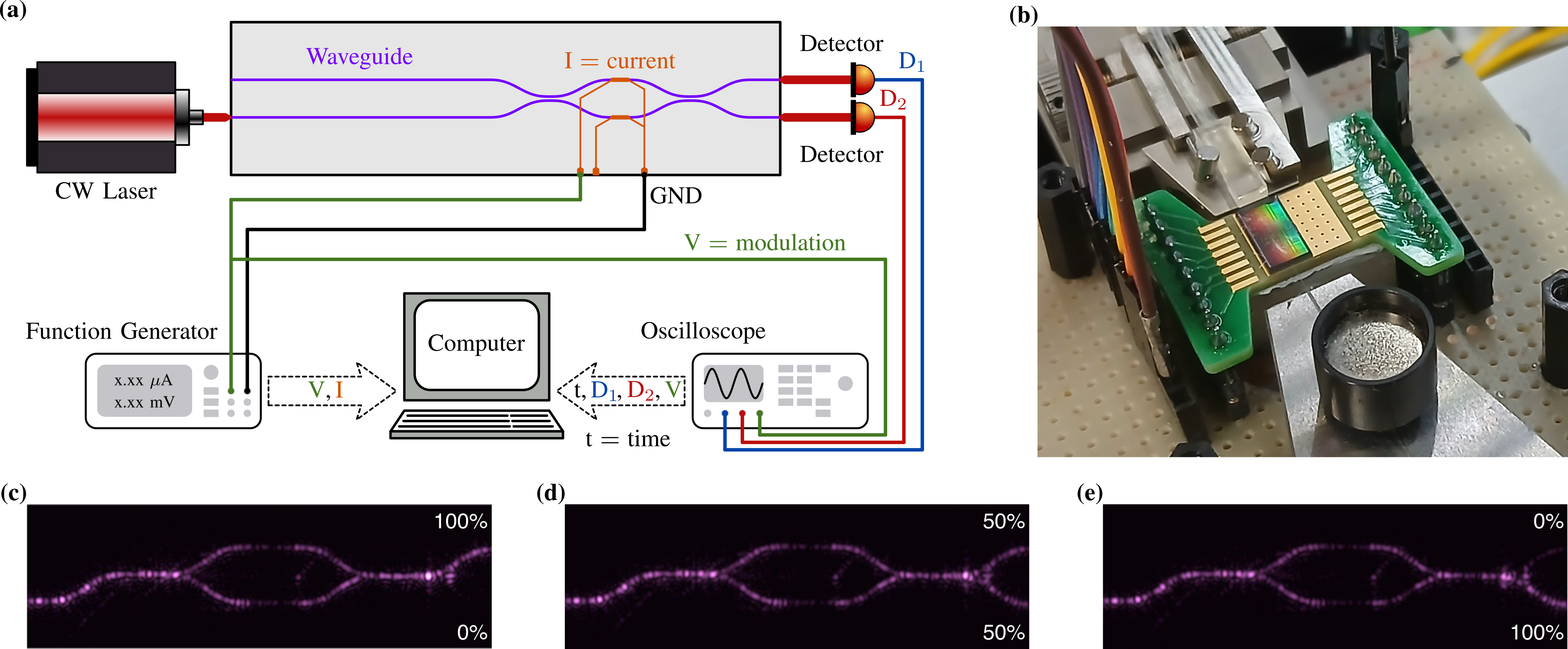}
    \caption{
    \textbf{Experimental setup for optoelectronic characterization of TiN heaters, conceptual diagram. (a)} A continuous-wave laser at either $\lambda = 810$~nm for Design~1 or $\lambda = 1550$~nm for Design~2 provides the optical input to the device under test. The electrical signal is generated by the function generator and driven into the chip heater. The oscilloscope measures in time-domain the modulation and the electrical signal of two photodetectors ($D_1$ and $D_2$) collecting light from both outputs of an MZI. A computer is connected via USB to both the function generator and the oscilloscope for data collection. \textbf{Photograph of the mounted device. (b)} The photonic chip is wire-bonded to a custom PCB to access the electrical pads to drive the heaters, while a lensed input fiber and a lensed fiber array were aligned to the optical input and output ports, respectively. \textbf{NIR camera images of the chip under illumination at 810~nm (c)--(e)} As the heater voltage is varied, the optical power distribution at the MZI outputs is controlled: (c) light predominantly routed to one output port, (d) approximately equal 50:50 splitting between the two output ports, and (e) light routed to the opposite output port.}
    \label{fig:Experimental_setup}
\end{figure*}

\begin{table}[h]
    \centering
    \caption{Safe operating limits extracted from the electrical breakdown characterization.}\label{tab:Operating_Conditions}
    \begin{tblr}{
                colspec = {X[40,c,m]X[50,c,m]X[60,c,m]X[50,c,m]X[50,c,m]},
                rowsep = 5pt,
                row{1} = {bg = TitleBlue, fg = white},
                row{2} = {bg = RowBlue},}
        \hline
        Design & $V_\mathrm{max}$~(\unit{\volt}) & $I_\mathrm{max}$~(\unit{\milli\ampere}) & $P_\mathrm{max}$~(\unit{\mW}) & $R_\mathrm{max}$~(\unit{\ohm}) \\
        \hline
        1 & 11.7~\textpm~0.3 & 15.3~\textpm~0.1 & 179~\textpm~3 & 760~\textpm~20 \\
        2 & 8.6~\textpm~0.3 & 15.9~\textpm~0.4 & 136~\textpm~2 & 540~\textpm~30 \\
        \hline
    \end{tblr}
\end{table}

\subsection{Electro-optical characterization}\label{sec:Electro_Optical_Characterization}

The electro-optical response of the heaters was investigated using the dedicated setup schematically depicted in Fig.~\ref{fig:Experimental_setup}a. Although the PICs are more complex, this schematic highlights only the MZI and its associated heaters, which are the primary focus of the present study. The electrical connection between the photonic chip and the driving electronics was established via an interposer and a custom-designed Printed Circuit Board (PCB). The chip was wire-bonded to this PCB, granting access to the targeted on-chip heaters. Optical measurements were performed by injecting a continuous-wave (CW) laser beam into one of the MZI input waveguides via butt-coupling with a lensed fiber, as shown in Fig.~\ref{fig:Experimental_setup}b. The laser wavelength was set to \qty{810}{\nm} for Design~1 and \qty{1550}{\nm} for Design~2. At the device output, light from both MZI ports was simultaneously collected by a lensed fiber array and routed to two independent photodetectors. The electrical outputs from these photodetectors, along with the modulation signal applied to the heater, were acquired in parallel across three channels of a digital oscilloscope. This configuration enabled simultaneous monitoring of the optical response at each MZI output as a function of the applied heater voltage. The redistribution of optical power between the MZI outputs as a function of the applied heater bias was visually captured using a NIR camera, providing a direct qualitative confirmation of the phase-shifting operation, as visible in Fig.~\ref{fig:Experimental_setup}c--e. Both the oscilloscope and the function generator were connected to a computer via USB to ensure synchronized control and automated data acquisition. Using this experimental setup, the heaters were characterized in two distinct operating regimes. In the DC regime, the function generator applied a slow controlled voltage ramp while simultaneously measuring the current, allowing for the extraction of the $\pi$-shift power as detailed in Section~\ref{sec:Pi_Shift_Power}. In the AC regime, the function generator drove the heaters with either sinusoidal or square-wave voltage waveforms. These waveforms were used to characterize the frequency response and the transient thermal dynamics, respectively, as discussed in Sections~\ref{sec:Frequency_Response} and~\ref{sec:Time_Response}. The electro-optical characterization was performed on 12 MZIs for Design~1 and 12 MZIs for Design~2. The reported uncertainties correspond to the standard deviation evaluated within the same Design.

\begin{figure*}[t]
\centering
\includegraphics[width = \linewidth]{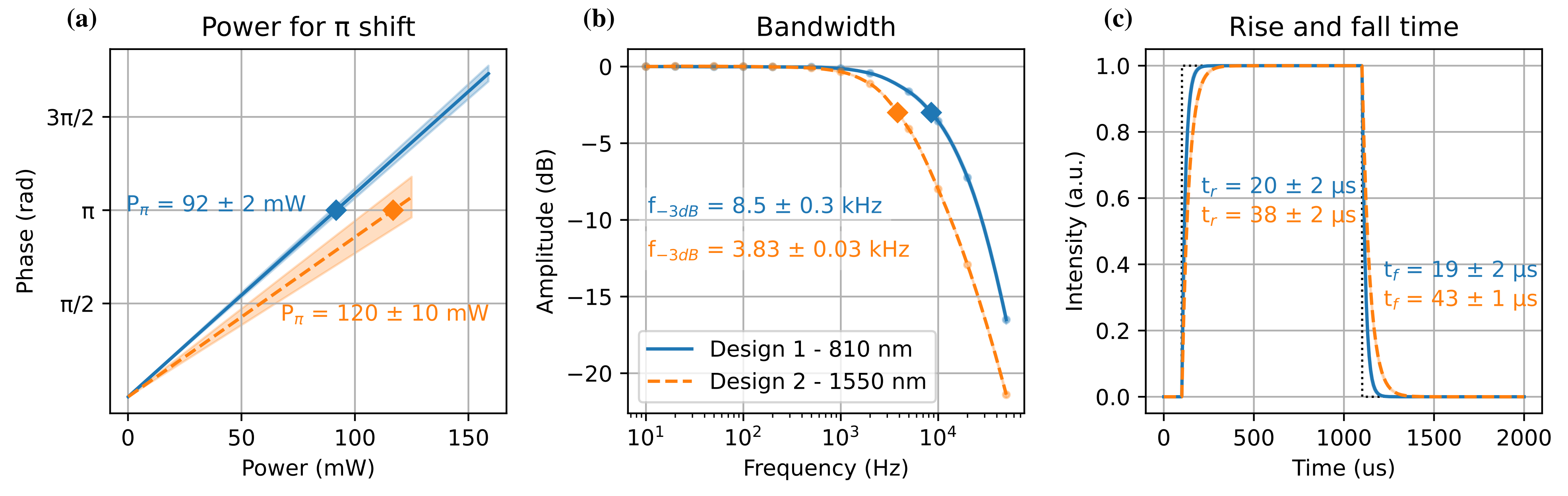}
\caption{\textbf{Electro-optic characterization of the thermo-optic phase shifters.}
\textbf{(a)} Phase shift as a function of dissipated electrical power, from which the $\pi$-shift power $P_{\pi}$ was extracted.
\textbf{(b)} Frequency response of the normalized optical modulation amplitude, from which the $-3$~dB cut-off frequency was extracted.
\textbf{(c)} Normalized transient response under square-wave excitation, from which the rise and fall time constants were obtained through exponential fitting.
Blue and orange curves refer to Design~1 and Design~2, respectively. Solid lines indicate mean values and shaded regions the corresponding standard deviation.}
\label{fig:Electro_Optic_Summary}
\end{figure*}

\subsubsection{\texorpdfstring{$\pi$}{pi}-shift power}\label{sec:Pi_Shift_Power}

To measure the $\pi$-shift power, the function generator was programmed to apply a double voltage ramp (from \qty{0}{\volt} to $V_\mathrm{max}$ and back to \qty{0}{\volt}), utilizing a current compliance limit of $I_\mathrm{max}$ based on the SOA defined in Sec.~\ref{sec:Heaters_Working_Range}. Since a double voltage ramp was used, the forward and backward sweeps were analyzed separately to check for hysteresis: no significant hysteresis was observed within the uncertainty of the measurement. 
The phase shift $\Delta\varphi$ induced by the heater was derived from the photodetector signals by inverting the MZI transfer function, which is described by a sinusoidal-squared relation\cite{Doughan2024Strip}. For each heater, the optical signals measured at the two output ports were fitted to the function $\alpha\sin^2(\beta P+\gamma)+\delta$, where $P$ represents the dissipated electrical power and $\alpha,\beta,\gamma,\delta$ are the fitting parameters. The fitted signal was subsequently normalized to eliminate the amplitude ($\alpha$) and offset ($\delta$) terms, and the corresponding phase was reconstructed using the inverse of the sinusoidal transfer function. The incremental phase variation was cumulatively summed over the entire power range. This unwrapping procedure generates a continuous, monotonic phase response that extends beyond the standard $[0,2\pi]$ interval, effectively capturing phase excursions larger than $2\pi$. The results are illustrated in Fig.~\ref{fig:Electro_Optic_Summary}a, displaying $\Delta\varphi$ as a function of applied electrical power for both designs. 
As anticipated for a thermo-optic phase-shifting mechanism\cite{Ribeiro2020Column}, the phase shift demonstrates an approximately linear dependence on the dissipated electrical power, with $\beta$ acting as the proportionality constant. The $\pi$-shift power $P_{\pi}$, defined as the electrical power necessary to accumulate a phase difference of $\pi$ between the two MZI arms, was ultimately determined via interpolation of the reconstructed phase curve. For each design, the phase--power curves from all analyzed devices were averaged, and the corresponding standard deviation was used to estimate the uncertainty of $P_{\pi}$. The extracted $P_{\pi}$ values are detailed in Tab.~\ref{tab:MeasurementsSummary}. Specifically, Design~1 yielded a $\pi$-shift power of $P_{\pi} = \qty{92 \pm 2}{\mW}$, while Design~2 required a higher power of $P_{\pi} = \qty{120 \pm 10}{\mW}$.
It is worth noting that, for Design~2, the extracted $P_{\pi}$ corresponds to
approximately 88\% of the SOA power limit, compared to about 51\% for
Design~1. This reduced operating margin highlights the need for further
power-efficiency optimization of the \qty{1550}{\nm} platform, for instance
through thermally isolated geometries such as substrate-undercut structures.

\subsubsection{Frequency response}\label{sec:Frequency_Response}

To evaluate the modulation bandwidth of the phase shifters, a small-signal sinusoidal modulation was superimposed onto a DC bias voltage applied to the heater. The DC bias was carefully selected to operate the MZI at quadrature (i.e., near a phase offset of $\pi/2$)\cite{Doughan2024Strip}, ensuring that the optical output varied approximately linearly with the applied power. A small AC perturbation, $V_\mathrm{AC}\sin(\omega t)$ where $\omega = 2\pi f$, was then added to this bias, resulting in a total driving signal:
\begin{equation}
    V(t) = V_\mathrm{DC} + V_\mathrm{AC}\sin(\omega t + \phi).
\label{eq_sin_bode_fit}
\end{equation}
Operating within this linear regime guarantees that the resulting optical modulation at the MZI output is sinusoidal in power, thereby simplifying the characterization of the frequency response. Measurements were conducted over a frequency range from \qty{10}{\Hz} to \qty{100}{\kHz}, utilizing logarithmically distributed sampling points (1, 2, and 5 points per decade) to accurately reconstruct the electro-optic frequency response of each device. The optical modulation amplitude was extracted via a sinusoidal fit using Eq.~\ref{eq_sin_bode_fit} as a function of the applied modulation frequency. The amplitude response of each channel was then normalized to the value measured at the lowest frequency point, $A_0$, and converted to decibels as $20\log_{10}(A/A_0)$. Because the responses of the two interferometric output channels differ by a $\pi$ phase shift, they were properly averaged to yield a single representative electro-optic transfer function per device. For both designs, the averaged amplitudes from all analyzed devices were combined to determine the mean frequency response and its standard deviation. A representative Bode plot is presented in Fig.~\ref{fig:Electro_Optic_Summary}b, plotting the normalized optical modulation amplitude against frequency for Design~1 and Design~2. The cut-off frequency $f_c$ was identified at the \qty{-3}{\dB} point of the mean response curve, with its uncertainty estimated from the upper and lower bounds of the standard-deviation band. The extracted cut-off frequencies for both designs are summarized in Tab.~\ref{tab:MeasurementsSummary}. Design~1 exhibited a significantly higher modulation bandwidth with a cut-off frequency of $f_c = \qty{8.5 \pm 0.3}{\kHz}$, compared to $f_c = \qty{3.83 \pm 0.03}{\kHz}$ for Design~2.

\subsubsection{Time response}\label{sec:Time_Response}

To complement the frequency-domain analysis, the temporal response of the phase shifters was investigated to provide an independent estimation of the thermal cut-off frequency. The function generator was programmed to deliver a \qty{10}{\Hz} square-wave signal, defined as $V(t) = V_\mathrm{DC} + V_\mathrm{AC}\operatorname{square}\left(\omega t\right)$. The $\operatorname{square}\left(\omega t\right)$ function oscillates between 0 and 1, effectively driving the heater between its off and on states with sharp, well-defined transition edges. The temporal dynamics of the thermo-optic phase shifters were analyzed by recording the transient optical signals at both MZI output ports under this square-wave excitation. For every device, the rising (off-to-on) and falling (on-to-off) transitions were evaluated independently. The exact switching instant was first identified from the electrical driving waveform; subsequently, the corresponding optical traces for both output channels were fitted separately. Each transient was modeled using a single-exponential response:

\begin{equation}
I(t)=A\left(1-e^{-\frac{t-t_0}{\tau}}\right)+D, \qquad t \ge t_0
\end{equation}

where $A$ represents the signal amplitude, $D$ is the offset term, $t_0$ is the estimated switching time, and $\tau$ denotes the characteristic time constant. Depending on the selected output port and the direction of the transition (rising or falling), the fitted amplitude $A$ could take either a positive or negative sign, consistent with the complementary nature of the MZI outputs. Distinct values for $A$, $D$, and $\tau$ were extracted for the rising and falling edges of each trace, offering a clear estimate of the phase shifter's thermal turn-on and turn-off dynamics. These fitting results, obtained independently for both output ports, were then aggregated across all analyzed devices to calculate the mean rise ($\tau_r$) and fall ($\tau_f$) time constants, along with their standard deviations. The resulting normalized transient responses are depicted in Fig.~\ref{fig:Electro_Optic_Summary}c, where the shaded regions indicate the variability of the extracted time constants. The measured rise and fall time constants are also documented in Tab.~\ref{tab:MeasurementsSummary}, alongside the cut-off frequencies derived from the Bode analysis, facilitating a direct comparison between the time-domain and frequency-domain assessments of the device performance. For Design~1, the transient constants were found to be $\tau_r = 20 \pm 2$~\unit{\us} and $\tau_f = 19 \pm 2$~\unit{\us}. For Design~2, the respective constants were longer, measuring $\tau_r = 38 \pm 2$~\unit{\us} and $\tau_f = 43 \pm 1$~\unit{\us}.

Discrepancies between $\tau_r$ and $\tau_f$ often arise in practical thermo-optic devices because the heating and cooling dynamics are not necessarily governed by identical boundary conditions. Specifically, the turn-on transient is actively driven by Joule heating within the resistor, whereas the turn-off transient relies entirely on passive heat dissipation into the surrounding materials and interfaces. Consequently, differing effective thermal paths and thermal resistances can lead to asymmetric transient constants for the rising and falling edges.

\begin{table}[h]
    \centering
    \caption{Summary of the electro-optical characterization results for the two investigated designs.}
    \label{tab:MeasurementsSummary}
    \begin{tblr}{
        colspec = {X[40,c,m] X[50,c,m] X[50,c,m] X[50,c,m] X[50,c,m]},
        rowsep = 5pt,
        row{1} = {bg = TitleBlue, fg = white},
        row{2} = {bg = RowBlue},
    }
        \hline
        Design & $P_{\pi}$~(\unit{\mW}) & $f_c$~(\unit{\kHz}) & $\tau_r$~(\unit{\us}) & $\tau_f$~(\unit{\us}) \\
        \hline
        1 & \num{92 \pm 2} & \num{8.5 \pm 0.3} & \num{20 \pm 2} & \num{19 \pm 2} \\
        2 & \num{120 \pm 10} & \num{3.83 \pm 0.03} & \num{38 \pm 2} & \num{43 \pm 1} \\
        \hline
    \end{tblr}
\end{table}

The results summarized in Tab.~\ref{tab:MeasurementsSummary} show that
Design~1 combines a lower $\pi$-phase-shift power
($P_{\pi}=\qty{92\pm2}{\mW}$) with a faster dynamic response. This is
reflected in its higher cut-off frequency ($f_c=\qty{8.5\pm0.3}{\kHz}$)
and shorter rise and fall times ($\tau_r=\qty{20\pm2}{\us}$ and
$\tau_f=\qty{19\pm2}{\us}$). By contrast, Design~2 requires a higher
electrical power to induce a $\pi$ phase shift and exhibits slower
thermal dynamics, as indicated by its lower modulation bandwidth and
longer transient response times.

These differences arise from the combined effect of the material stack,
heater geometry, and operating wavelength. In a thermo-optic phase
shifter, the accumulated phase shift scales as
$\Delta\Phi=(2\pi L/\lambda_0)(dn_{\mathrm{eff}}/dT)\Delta T$ \cite{harris_efficient_2014}; therefore, for otherwise identical devices, operation at a longer wavelength requires a larger temperature-induced effective-index change, or equivalently a
longer interaction length, to obtain the same phase shift. Since the heater length is the same in the two designs, this wavelength scaling contributes to the larger $P_{\pi}$ observed for Design~2. In addition, the \qty{1550}{\nm} implementation requires a thicker oxide stack to suppress substrate leakage and maintain adequate optical isolation. This increases the thermal separation between the heater and the optical mode and raises the effective thermal inertia, leading to a higher $\pi$-shift power and a slower temporal response. Conversely, the thinner oxide stack used in Design~1 improves thermal coupling to the waveguide and reduces the heated thermal volume, yielding a more efficient and faster thermo-optic response.

The results in Tab.~\ref{tab:MeasurementsSummary} place the proposed heaters among competitive non-suspended SiN thermo-optic phase shifters, especially when considering the combined metrics of $P_{\pi}$, kHz-range modulation bandwidth, and process simplicity \cite{gao_comprehensive_2023}. While suspended and undercut architectures can achieve substantially lower switching powers, they typically introduce additional processing complexity and a stronger trade-off with thermal response time.

\section{Conclusion}\label{sec:Conclusion}
In this work, we demonstrated the design, fabrication, and characterization of state-of-the-art TiN-based phase shifters integrated onto SiN waveguides. Our approach introduces a CMOS-compatible fabrication workflow that uses a single lithographic step to simultaneously define high-resistivity TiN heaters and
low-resistance Al interconnects. By eliminating the need for complex
post-fabrication processing or multi-step metal definition, our approach significantly simplifies the manufacturing process of reconfigurable PICs while maintaining stable operation within the exploreded operating range.

Our experimental results establish these devices as a competitive solution across different spectral regimes. At \qty{810}{\nm}, we achieved a $\pi$-shift power ($P_{\pi}$) of \qty{92\pm2}{\mW} and a \qty{3}{dB} modulation bandwidth of \qty{8.5\pm0.3}{\kHz}. We also identified a safe operating area, confirming that these performance metrics can be maintained within reliable thermal and electrical limits. Although the \qty{1550}{\nm} platform exhibits higher thermal inertia, mainly due to the thicker cladding required for optical isolation, it remains a robust and scalable solution for C-band PIC applications. Overall, these results demonstrate that the Al--Ti--TiN--Ti multilayer architecture provides an efficient and accessible platform for applications ranging from quantum information processing to microwave photonics and biosensing.

Future work will focus on improving the power efficiency of the
\qty{1550}{\nm} platform. Substrate undercut could substantially reduce the thermal mass and heat leakage, potentially lowering the $\pi$-shift power by up to an order of magnitude, although likely at the cost of a slower thermal response. The optimal design will therefore depend on the target application, balancing speed, power consumption, thermal crosstalk, and robustness.

\section*{Acknowledgment}

This research has been supported in part by the Autonomous Province of Trento, L.P. 6 luglio 2023, n. 6 e ss.mm.ii., project FANES Cup: C69J24000350001, by the PNRR MUR project PE0000023-NQSTI (Italy), by the European Union’s Horizon 2020 Research and Innovation Programme under Grant No. 899368 – EPIQUS and by the European Union’s Horizon JU Research and Innovation Actions under Grant No. 101213727 - PIXEurope.

\bibliographystyle{IEEEtran}
\bibliography{Bibliography}
\end{document}